\documentclass[aps,prc,showpacs,preprint]{revtex4}
\usepackage{graphicx}
\usepackage[dvips]{color}
\usepackage{colordvi}
\usepackage{bm}
\begin{document}
\bibliographystyle{apsrev}

\title{In medium $T$-matrix for nuclear matter with three-body forces -
  binding energy and single particle properties}\thanks{Research supported in
  part by the Polish Ministry of Science and Higher Education, grant N N202 1022 33}
\author{     V. Som\`a\footnote{Electronic address~:
vittorio.soma@ifj.edu.pl}}
\affiliation{
Institute of Nuclear Physics PAN, PL-31-342 Krak\'ow, Poland}
\author{     P. Bo\.{z}ek\footnote{Electronic address~:
piotr.bozek@ifj.edu.pl}}
\affiliation{
Institute of Nuclear Physics PAN, PL-31-342 Krak\'ow, Poland\\
and\\
Institute of Physics, Rzeszow University, PL-35-959 Rzesz\'ow, Poland}

\date{\today}

\begin{abstract}
We present  spectral calculations of nuclear matter
 properties including three-body forces. Within the in-medium $T$-matrix 
approach,
implemented with the 
CD-Bonn and Nijmegen  \cite{cdbonn, nijmegen} 
potentials plus the  three-nucleon Urbana interaction \cite{Carlson:1983}, we
compute the energy per particle in symmetric and neutron matter.
 The three-body forces are included via an effective density 
dependent two-body force in the
in-medium  $T$-matrix equations. After fine tuning the parameters of
 the three-body force to reproduce the phenomenological saturation 
point in symmetric nuclear matter, we calculate the incompressibility and
 the energy per particle in neutron matter. We find a soft equation of state 
 in symmetric nuclear matter but a relatively large value of the symmetry 
energy. We study the 
the influence of the three-body forces on the single-particle properties.
For symmetric matter the spectral function is 
 broadened at all momenta and all
densities, while an opposite effect is found for the case of neutrons only.
Noticeable modification of the spectral functions are realized only 
for densities
 above the saturation density.
The modifications of the self-energy and the  effective mass are not
very large and appear to be strongly suppressed above the Fermi momentum.
\end{abstract}


\maketitle

\section{Introduction}
\label{sec:intro}

Many-body calculations based on \textit{realistic} bare nucleon-nucleon
potentials are able to reproduce qualitatively but not quantitatively the
saturation properties of symmetric nuclear matter. The saturation points 
calculated  with different
 approaches implemented with
various choices of the nucleon-nucleon (NN) potential happen to lie 
  on the so called Coester line in the energy per particle-density plane,
 away from the experimentally allowed values \cite{Coester:1970ai}. 
The theoretical predictions give a 
saturation density sensibly higher than the experimental value 
$\rho_0 \approx 0.16 \, \mbox{fm}^{-3}$ (usually in the range 
$1.5 \, \rho_0 - 2 \, \rho_0$) and
often overbind the nuclear system (up to $25\%$), failing to 
get close to the
empirical binding energy $E_0 \approx -16 \, \mbox{MeV}$. These
 discrepancies and uncertainties get amplified when calculating 
the equation
of state (EOS) of pure neutron matter, which is necessary for the estimates
 of key quantities
such as the symmetry energy and in general for the description
of neutron-rich matter in neutron stars.

It is believed that these deficiencies originate from neglecting 
three-body forces (TBF)
between nucleons in the dense medium. 
Variational and Brueckner-Hartree-Fock (BHF) calculations
which take TBF into account achieve a realistic description 
of cold nuclear
matter, reproducing the saturation point of the symmetric 
EOS with a good degree of
accuracy \cite{Akmal:1998, Wiringa:1988tp, vcs1, fabrocini1, fabrocini,
  baldonstar, lombardonstar, EOSBHF, Baldo:2007}. 

The self-consistent real-time  Green's functions technique can be used 
practically to perform calculations of in-medium propagators and properties of 
 a many-body system. In nuclear matter this approach
considers the in-medium $T$-matrix approximation for
 the two-particle Green's functions. This approximation scheme
 incorporates the  resummation of the diagrams necessary in the presence 
of a strong repulsive short range component in nuclear forces and 
 at the same time is thermodynamically consistent 
\cite{KadanoffBaym, Bozek:2001tz}.
In the zero temperature limit, the in-medium $T$-matrix method 
 gives results similar to the variational and
BHF approaches \cite{Bozek:2002tz, Bozek:2002ry, Dewulf:2003nj, Rios:2007gp}
 for the binding
 energy. 
At finite temperatures, however, it has the important advantage
of automatically fulfilling the thermodynamic consistency relations,
 guaranteeing
an unambiguous estimate of the thermodynamic quantities \cite{Bozek:2001tz}.
It has been employed in several works including studies of the modification of
nucleon properties in the medium \cite{di1, Bozek:2002em}, effective scattering 
\cite{Dickhoff:1999yi},
pairing correlations \cite{Bozek:2001nx, Bozek:2002jw}, liquid-gas
phase transition \cite{Rios:2008ef} and nuclear matter binding and thermodynamics
\cite{Bozek:1998su, Dewulf:2002gi, Bozek:2002tz, Bozek:2002mw, Dewulf:2003nj,
  Frick:2003sd, Soma:2006, Rios:2007gp}.

Existing self-consistent $T$-matrix  calculations take into account only 
the  two-body interactions, with the exception of  a first attempt using
a  purely
phenomenological Lagaris-Pandharipande TBF \cite{Soma:2007ew}. 
In the present work we
implement the self consistent $T$-matrix scheme with two different realistic
NN potentials (CD-Bonn and Nijmegen) plus Urbana three-nucleon forces,
which consist in an attractive part, based on the two-pion exchange,
 and a repulsive
term related to the presence of a Roper resonance as an intermediate state.

Within this approach we calculate the binding energy per particle 
for both symmetric and
pure neutron matter, and give an estimate of the nuclear symmetry energy.
Moreover we study the nucleon spectral function, self-energies and effective
mass in comparison with the case of two-body forces only.

The paper is organized as follows. The next section is  a
brief review of the
Green's function formalism and the $T$-matrix approximation. In the third section we
illustrate how the three-body forces are included in the $T$-matrix scheme, and 
in section IV we present our results on the nuclear EOS and the single
particle properties, placing emphasis on the modifications due to
the introduction of TBF. In the last section we end with a discussion and a
short summary.

At the moment we have considered the zero temperature case only.  We plan to
perform calculations at finite temperature, yielding  thermodynamic quantities
and  a more complete nuclear EOS including TBF. 
This will be the object of a future 
publication.

\section{Self-consistent $T$-matrix approach}
\label{sec:tmatrix}

In the diagrammatic expansion of the nucleon self-energy, the  summation
of the \textit{ladder} diagrams at all orders leads to the $T$-matrix or ladder
approximation \cite{KadanoffBaym, blarip}.
The imaginary part of the self-energy is obtained as follows (for more details see
\cite{Soma:2006} and references therein)
\begin{eqnarray}
\label{eq:imsigma}
\mbox{Im}\,\Sigma^{R}(\mathbf{p},\omega)&=&\int   \frac{d^3{k}}{(2\pi)^3}
\frac{d{\omega^{'}}}{2\pi}\left[ \mbox{Im}\, 
\langle(\mathbf{p-k})/2|T^{R}(\mathbf{p+k},\omega+\omega')|(\mathbf{p-k})/2\rangle
\right. \\ & &\left. \nonumber 
- \mbox{Im}\, \langle(\mathbf{p-k})/2|T^{R}(\mathbf{p+k},\omega+\omega')|
(\mathbf{k-p})/2\rangle
\right]\left[
b(\omega+\omega^{'})+f(\omega^{'})\right]A(\mathbf{k},\omega^{'}) \, .
\end{eqnarray}
Here $b(\omega)$ and $f(\omega)$ are respectively the Bose-Einstein and the
Fermi-Dirac 
distributions, $A(\mathbf{p},\omega)$ is the nucleon spectral function
related to the single particle propagator through
\begin{equation}
\label{eq:gsmaller}
G^< (\mathbf{p},\omega) = f(\omega) \, A(\mathbf{p},\omega) \, .
\end{equation}
The self-consistent in-medium two-particle scattering matrix T is defined as
\cite{Danielewicz:1982kk}
\begin{eqnarray}
\label{eq:tmatrix}
& \langle\mathbf{k}|T^{R(A)}(\mathbf{P},\Omega)|\mathbf{k'}\rangle = 
V(\mathbf{k},\mathbf{k'})
\nonumber \\
& + \displaystyle \int \frac{d^3{p}}{(2\pi)^3} 
\frac{d^3{q}}{(2\pi)^3}
V(\mathbf{k},\mathbf{p}) 
\langle\mathbf{p}|{G_2^{nc\ R(A)}}(\mathbf{P},\Omega)|\mathbf{q}\rangle
\langle\mathbf{q}|T^{R(A)}(\mathbf{P},\Omega)|\mathbf{k'}\rangle \: ,
\end{eqnarray}
where $V(\mathbf{k},\mathbf{k'})$ is the chosen nucleon-nucleon potential and
the uncorrelated two-particle Green's function $G_2^{nc}$ is the 
product of two dressed one-body propagators
\begin{eqnarray}
\label{eq:g2nc}
& \langle\mathbf{k'}|{G_2^{nc \ <(>)}}(\mathbf{P},\Omega)|\mathbf{k}\rangle =
\nonumber \\
& \displaystyle
i (2\pi)^3 \, \delta (\mathbf{k}-\mathbf{k'}) \int \frac{d\omega'}{2\pi}
G^{<(>)}(\mathbf{P}/2+\mathbf{k},\Omega-\omega') \,
G^{<(>)}(\mathbf{P}/2-\mathbf{k},\omega') \: .
\end{eqnarray}
The single particle propagators are dressed via the Dyson equation
\begin{equation}
\label{eq:dyson}
G^{R(A)\ -1}(\mathbf{p},\omega)=\omega-\frac{p^2}{2m}-
\Sigma^{R(A)}(\mathbf{p},\omega) \ .
\end{equation}
Formulae (\ref{eq:imsigma}), (\ref{eq:tmatrix}) and (\ref{eq:dyson})
represent a closed set of equations which has to be solved with an
iterative algorithm, until self-consistency is achieved. 
During the iteration the Fermi energy $\mu$ is adjusted to get the desired
density $\rho$ from
\begin{equation}
\rho= \int \frac{d^3{k}}{(2\pi)^3}
\frac{d{\omega}}{2\pi} G^< (\mathbf{p},\omega)\ \  .
\end{equation}
Once stable solutions
are obtained, the energy of the many-body system is calculated directly from
the expectation value of the Hamiltonian (see \cite{Soma:2006}), avoiding the
use of the Galitskii-Koltun sum rule. This is an important point because the
sum rule, besides showing some dependence on the chosen energy integration
cutoff, loses its validity in the presence of three-body forces.

\section{Three-body forces}
\label{sec:tbf}

The profound reason for the appearance of three-nucleon forces is that protons
and neutrons are not unstructured particles, as  they are considered
when constructing the various NN potentials in meson-nucleon field theory.
When a nucleon interacts its internal structure indeed can be modified; 
this implies that there are interactions
which are not anymore simply of the nucleon-nucleon type.
This is usually described via the excitation of one nucleon to another state
or resonance, giving rise to processes
which cannot be decomposed into a sequence of NN diagrams.
At low densities their contribution is negligible, while they become relevant
at densities involved in nuclear matter calculations \cite{Grange:1989}.

All calculations aiming at realistic predictions of nuclear matter properties
around and above the saturation point include TBF. 
There exist few works which implement TBF in nuclear calculations and test
their effects 
on binding energy, pairing and single-particle properties
 in different approaches, both in finite systems and in nuclear matter 
\cite{urbanatbf, PhysRevC.56.1720, Zuo:2002sg, Zhou:2004br, Zuo:2002sfa,
  Zuo:2006nz, Baldo:2007, lombardonstar, EOSBHF}.

We consider here the  approach of the Urbana three-nucleon potential
developed by Carlson
\textit{et al} \cite{Carlson:1983}. It is composed of two terms
\begin{equation}
\label{eq:tbf-general}
V_{ijk} = V_{ijk}^{2\pi} + V_{ijk}^{R}.
\end{equation}
The first part is constructed from two-pion exchange, where a $\Delta$
resonance appears as intermediate state.
Its contribution is attractive and dominates at low densities
(below nuclear saturation).
The second term, whose nature is more phenomenological, provides the repulsion
needed at higher densities. It is introduced to represent the contribution of
several other diagrams, possibly including in an effective way also four-
and more nucleon interactions. Alternative implementations and microscopic
derivations of TBF are also available \cite{Grange:1989}.

In more details, the Urbana TBF have the following structure
\begin{equation}
\label{eq:3b-2pi}
V_{ijk}^{2\pi} = A \sum_{cyc}
\left (
\left \{
X_{ij},X_{jk}
\right \}
\left \{
\bm{\tau}_i \cdot \bm{\tau}_j, \bm{\tau}_j \cdot \bm{\tau}_k
\right \}
+ \frac{1}{4}
\left [
X_{ij},X_{jk}
\right ]
\left [
\bm{\tau}_i \cdot \bm{\tau}_j,\bm{\tau}_j \cdot \bm{\tau}_k
\right ]
\right) \: ,
\end{equation}
where 
\begin{equation}
X_{ij} = Y(r_{ij}) \: \bm{\sigma}_i \cdot \bm{\sigma}_j + T(r_{ij}) \: S_{ij} \: ,
\end{equation}
and
\begin{equation}
\label{eq:3b-r}
V_{ijk}^{R} = U \sum_{cyc} T^2(r_{ij}) \: T^2(r_{jk}) \: .
\end{equation}
The radial functions $Y(r)$ and $T(r)$ are the Yukawa and tensor functions
respectively, the tensor operator is defined as
$S_{ij} = 
3 \, (\bm{\sigma}_i \cdot \bm{\hat{r}}_{ij})(\bm{\sigma}_j \cdot \bm{\hat{r}}_{ij}) 
- \bm{\sigma}_i \cdot \bm{\sigma}_j$, where $\bm{\hat{r}}_{ij}$ is the unit
vector of the distance between particles $i$ and $j$.
To determine the overall strength 
of the TBF and the relative strength between the two terms
two parameters are present ($A<0$ and $U>0$), to be tuned to reproduce the
saturation properties of symmetric nuclear matter. Since different NN
potentials lead to different saturation curves one should expect these
parameters to depend on the particular choice of the two-body force. 

The three-body interaction depends on the spatial, spin and isospin
coordinates of the three nucleons, and in such a form cannot be used in the
calculations. 
We then need to introduce some approximation and derive an effective
two-particle potential. This can be done by
averaging the action of the third nucleon, resulting in a mean field felt
by the other two:
\begin{equation}
V^3_{eff}(\bm{q},\bm{q'}) = \sum_{\sigma \, \tau}
\int \frac{d^3 k}{(2 \pi)^3} \: n(\bm{k}) \,
V^3(\bm{k},\bm{q},\bm{q'}) \: ,
\end{equation}
where $V^3(\bm{k},\bm{q},\bm{q'})$ is the Fourier transformed form of 
(\ref{eq:tbf-general}) and
\begin{equation}
n(\bm{k}) = \int \frac{d \omega}{2 \pi} \, G^< (\bm{k},\omega)
\end{equation}
is the particle momentum distribution. 
The sum over spin and isospin degrees of freedom just reminds us that
$V^3$ has a non
trivial structure in the $\sigma$ and $\tau$ spaces which has to be taken care of
(we didn't write explicitly spin and isospin indices).

This average has to be performed for each of the three nucleons and over all
their possible permutations, resulting in nine different terms. One has to pay
particular attention to the spin-isospin and tensor dependence of the various
averages and finally get, for each of the nine permutations, an effective
potential of the form
\begin{eqnarray}
V^3_{eff}(\bm{q},\bm{q'}) 
& = & V^{R}_s(\bm{q},\bm{q'}) + V^{2 \pi}_s(\bm{q},\bm{q'})
\\ \displaystyle
& + & V^{2 \pi}_{\sigma \tau} (\bm{q},\bm{q'}) \,
\bm{\sigma} \cdot \bm{\sigma'} \, \bm{\tau} \cdot \bm{\tau'} 
+ V^{2 \pi}_{S \tau}(\bm{q},\bm{q'}) \, S(\bm{q},\bm{q'}) \, 
\bm{\tau} \cdot \bm{\tau'} 
\: , \nonumber
\end{eqnarray}
where $V^{R}_s, V^{2 \pi}_s, V^{2 \pi}_{\sigma \tau}$ and $V^{2 \pi}_{S \tau}$
are now scalar functions.

Once we have obtained $V^3_{eff}$ (density dependent) we add it to the two-body potential in
(\ref{eq:tmatrix})
\begin{equation}
V \longrightarrow V' = V + V^3_{eff} \: .
\end{equation}
and perform the $T$-matrix iteration.


\section{Binding energy and single particle properties}
\label{sec:results}

We perform calculations with two different parameterizations of the NN
interaction, the CD-Bonn  and the Nijmegen potentials. 
For both of them we 
compute the energy per particle directly from the expectation value of the
interaction Hamiltonian, for symmetric and for pure neutron matter, with and
without TBF. 
In the case of three-body forces we have tuned the parameters $A$ 
and $U$ in (\ref{eq:3b-2pi}) and (\ref{eq:3b-r}) in the symmetric case
in order to reproduce the 
saturation density $\rho_0$ and binding energy $E_0$. Since the 
averaging over the third nucleon in TBF terms represents a rather crude
 approximation, the resulting numerical values of the parameters of
 the TBF are different than in other approaches.

\subsection{Symmetric nuclear matter}
\label{ssec:symm}

\begin{figure}[h]
\begin{center}
\includegraphics[width=8cm]{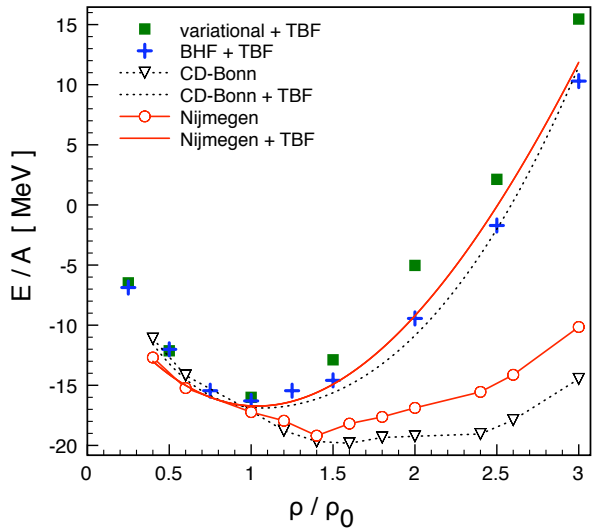}
\caption{(Color online) Energy per particle in symmetric nuclear matter as a function of
density (in units of the nuclear saturation density $\rho_0 = 0.16
 \ \mbox{fm}^{-3}$). 
$T$-matrix calculations are compared to the
variational \cite{Akmal:1998} and BHF \cite{Baldo:2007} approaches, both
including TBF.
}
\label{fig:en-symm}
\end{center}
\end{figure}

The energy per particle as a function of density for symmetric nuclear
matter is shown in Fig. \ref{fig:en-symm}. The calculations with only two-body
forces fail to reproduce the correct saturation behavior, predicting
a saturation density $\rho= 1.47 \, \rho_0$ in the case of the Nijmegen potential and
$\rho= 1.79 \, \rho_0$ for CD-Bonn. After the inclusion of three-nucleon interactions
the situation is significantly improved, with both curves saturating around
the phenomenological value $\rho_0=0.16$~fm$^{-3}$
and yielding a 
correct binding energy \footnote{We estimate the numerical error on all the
  energy calculations to be $\pm 0.5$ MeV, for details see \cite{Soma:2006}}
(Nijmegen $E_B=-16.4$ MeV and CD-Bonn $E_B=-16.3$ MeV). 
We notice that the effects of TBF are almost negligible at low densities and
become more and more significant as the density increases.

We plot for comparison also the results obtained with the variational 
method by Akmal et al. \cite{Akmal:1998} and with the BHF approach by Baldo
and Maieron \cite{Baldo:2007}, both using the Argonne v18 potential.
The overall agreement between all the
calculations is good. 

When we compute, however, the incompressibility at saturation
\begin{equation}
K_0 = 9 \left( \rho \, \frac{\partial^2 E}{\partial \rho^2} \right)_{\rho=\rho_0}
\end{equation}
we get for both potentials $K_0 \sim 150$ MeV, a
value smaller than the usual estimates $K_0 = 210 \pm 30$ MeV 
extracted from the  experimental data and alternative calculation schemes. 

\begin{figure}[h]
\begin{center}
\includegraphics[width=8cm]{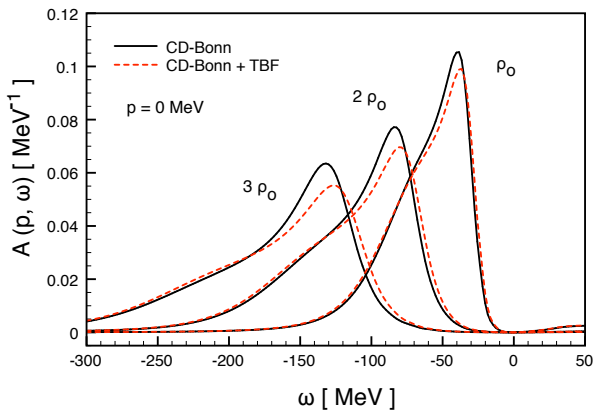}
\caption{(Color online) Spectral function at zero momentum 
for CD-Bonn interaction and  symmetric nuclear matter,
at $\rho_0$, $2 \, \rho_0$ and $3 \, \rho_0$.
}
\label{fig:spe-bs3-0}
\end{center}
\end{figure}

The structure (\ref{eq:gsmaller}) of the one-particle propagator allows us to
investigate the correlations induced in the dense medium by NN and three-body
forces. These correlations are reflected in the energy dependence of the
spectral function $A(p,\omega)$, which in the presence of the short-range
nuclear interactions differs substantially from the $\delta$-function
behavior that characterizes the quasiparticle approximation. In Fig. 
\ref{fig:spe-bs3-0} we show the spectral functions when only two-body forces
are present and when TBF are added for the CD-Bonn potential at three
different densities and zero momentum.
\begin{figure}[h]
\begin{center}
\includegraphics[width=12cm]{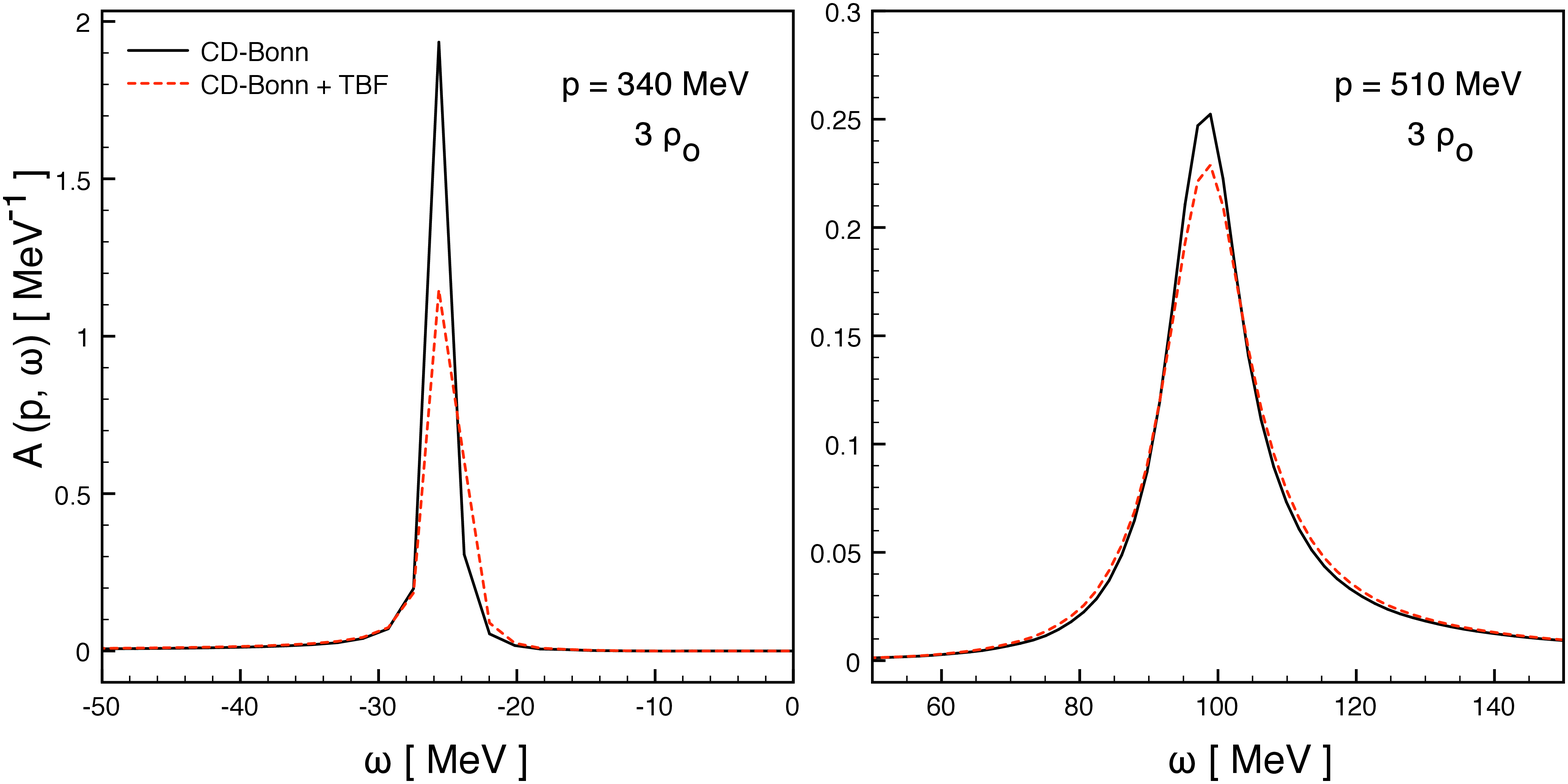}
\caption{(Color online) Spectral function at $p=340$ MeV (left panel) and
  $p=510$ MeV 
(right panel) 
for CD-Bonn interaction in symmetric nuclear matter.
}
\label{fig:spe-bs3-comb}
\end{center}
\end{figure}

The inclusion of TBF results in a further broadening of the peak of the
spectral function, with a larger effect at high densities. 
This behavior reflects an increase of the scattering between the nucleons and
it is common to all densities and all momenta. In Fig. \ref{fig:spe-bs3-comb} 
two cases with
non-zero momenta are presented at the density $3 \, \rho_0$. In the left panel
($p$ close to the Fermi momentum $p_F$),  the strongest reduction
of the peak shows up. The modifications caused by TBF are then slowly
disappearing for higher momenta, as illustrated in the right panel. The effect
of the TBF on the spectral function is minor for $\rho<\rho_0$. Consequently,
TBF and their isospin dependence cannot explain the strong dependence of the 
experimentally extracted proton spectral functions \cite{Rohe, :2007gqa,
  Bozek:2003wh, Hassaneen:2004ri, Konrad:2005qm} on the target nucleus.

\begin{figure}[h]
\begin{center}
\includegraphics[width=12cm]{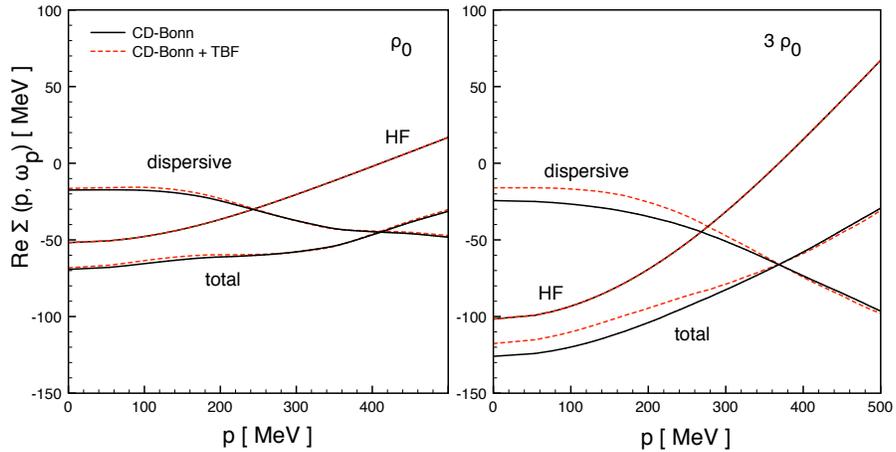}
\caption{(Color online) Nucleon self-energy for CD-Bonn interaction in
  symmetric nuclear matter  at $\rho_0$ (left panel) and $3 \, \rho_0$
(right panel)
}
\label{fig:self-bs-comb}
\end{center}
\end{figure}

We have analyzed the real part of the self-energy, which determines the
shift with respect to the free dispersion relation 
\begin{equation}
\omega_p = \frac{p^2}{2m} + \mbox{Re} \, \Sigma (\bm{p}, \omega_p) \: .
\end{equation}
It is the sum of the Hartree-Fock self-energy and a dispersive contribution
obtained via the imaginary part (\ref{eq:imsigma})
\begin{equation}
\label{eq:re-sigma}
\mbox{Re} \, \Sigma (\bm{p}, \omega) = \Sigma_{HF} (\bm{p}) +
\mathcal{P} \int \frac{d \omega'}{\pi} 
\frac{\mbox{Im} \, \Sigma^{R} (\bm{p}, \omega') }{\omega-\omega'} \: . 
\end{equation}
The introduction of TBF does not have
a big impact on the self-energy. As shown in Fig. \ref{fig:self-bs-comb} 
(left panel) for low densities the modification is smaller than $5 \%$, 
and it grows up to $10 \%$ at $\rho=3 \, \rho_0$ (right panel). At all densities
the effect is relevant for low momenta and basically vanishes above the 
Fermi surface. As expected TBF have a repulsive contribution, shifting
the dispersive part of $\mbox{Re} \, \Sigma (\bm{p}, \omega)$ 
up to less negative values.
The behavior is completely analogous if the Nijmegen potential is employed.

\begin{figure}[h]
\begin{center}
\includegraphics[width=12cm]{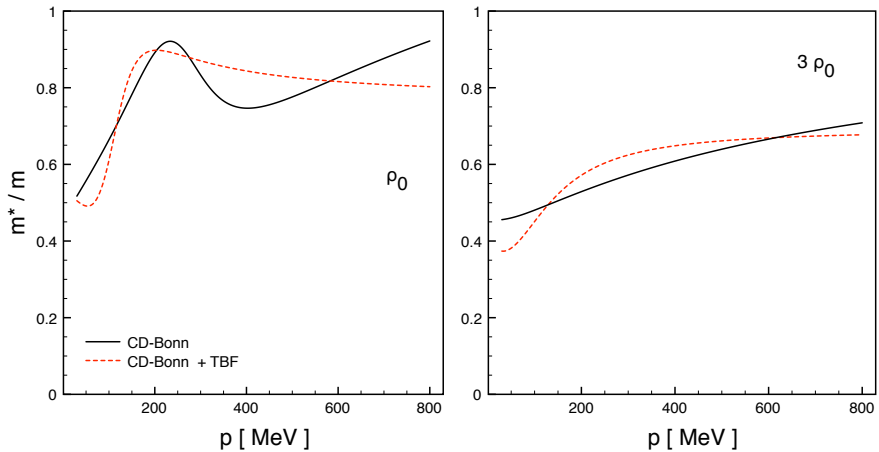}
\caption{(Color online) Effective mass (in units of the nucleon mass $m=939$ MeV) for the 
CD-Bonn interaction and symmetric nuclear matter 
 at $\rho_0$ (left panel) and $3 \, \rho_0$ (right panel).
}
\label{fig:effm-bs}
\end{center}
\end{figure}

Finally, the effective mass is evaluated in the presence of TBF.
 In general it is obtained at each momentum from
\begin{equation}
\frac{\partial {\omega}_p}{\partial p^2}
= \frac{1}{2 m^\star} \: .
\end{equation}

 The effective mass as function of the momentum  is displayed in Fig. \ref{fig:effm-bs}
for the CD-Bonn potential at two different densities. As for the self-energy,
the effect of the TBF  is not dramatic. At $\rho_0$ 
TBF smooth out the peak formed around
Fermi momentum $p_F = 263$ MeV, but do not change the value at the Fermi 
surface
$m^* \approx 0.85 \, m$ which is in agreement with the typical estimates from
 experiments \cite{Onsi:2002qf} and BHF calculation with rearrangement 
terms included.
Both with and without TBF the effective mass
at $3 \, \rho_0$ gets substantially lower, with   $m^* \approx 0.6 \, m$
at the Fermi momentum.
Similar behaviors are observed for the Nijmegen NN interaction with and without
three-body forces: at $\rho_0$ we found $m^* \approx 0.9 \, m$ and 
at $3 \, \rho_0$ we estimate $m^* \approx 0.65 \, m$ .

\subsection{Neutron matter}
\label{ssec:neut}

\begin{figure}[h]
\begin{center}
\includegraphics[width=8cm]{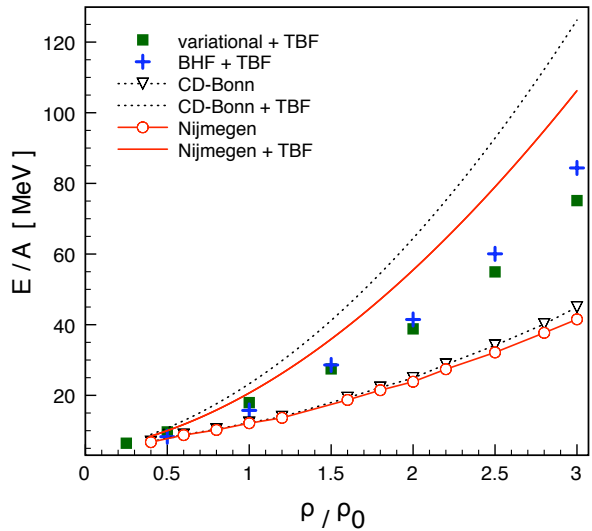}
\caption{(Color online) Energy per particle in neutron matter as a function of
density (in units of the nuclear saturation density $\rho_0 = 0.16 \ 
 \mbox{fm}^{-3}$) for
different potentials. 
The $T$-matrix calculations are compared to the
variational \cite{Akmal:1998} and BHF \cite{Baldo:2007} approaches, both
including TBF.
}
\label{fig:en-neut}
\end{center}
\end{figure}

In Fig. \ref{fig:en-neut} we show the results for the energy per particle in
the case of a system composed of neutrons only. For the two choices of the NN
potential, curves without and with TBF are displayed. When TBF are used, the
parameters $A$ and $U$ are fixed by
 the calculation in the  isospin symmetric case.
The introduction of TBF does not change qualitatively the
density dependence of $E/A$ but makes the EOS more stiff. As in the symmetric
case the results are compared to variational and BHF calculations
(both including TBF). We find  agreement at low densities, with
the $T$-matrix results getting stiffer at very high densities. A similar 
effect, a stiffer result from the $T$-matrix than from the BHF or
 variational calculations in neutron matter has been noticed when
 using two-body forces only \cite{Bozek:2002ry}.

An important quantity in determining the equation of state of 
isospin asymmetric nuclear matter is the symmetry energy. In
particular its
behavior is crucial for neutron star physics, since it affects the stability of
neutron star matter and controls the position of the crust-core transition
\cite{Kubis:2006kb}.
The symmetry energy is defined from the energy per nucleon $E/A$ as follows
\begin{equation}
E_{sym} (\rho) = \frac{1}{2} \left( \frac{\partial^2 E/A}{\partial \delta^2} 
\right)_{\delta=0} \: ,
\end{equation}
where $\delta=(\rho_n-\rho_p)/\rho$ is the so called asymmetry parameter.
We calculate $E_{sym}$ via the parabolic approximation, which is usually
employed to estimate the energy per particle for arbitrary asymmetries
(whose reliability has been checked  by Bombaci and Lombardo
\cite{Bombaci:1991zz})
\begin{equation}
\frac{E}{A} (n, \delta) = \frac{E}{A} (n, \delta = 0)
+  \delta^2 \, E_{sym} (n) \: .
\end{equation}
The experimental measurement of the symmetry energy at saturation 
$E_{sym} (\rho_0) \equiv S_0$ 
is not simple and still has a rather large uncertainty.
A common estimate is $S_0 = 32 \pm 6$ MeV \cite{Haensel:NS}. 
The density dependence of the symmetry energy is  of crucial importance for
neutron star physics. It  is fairly unknown, with large
differences, specially at high densities, between the various models
\cite{Klahn:2006ir}.

\begin{figure}[h]
\begin{center}
\includegraphics[width=8cm]{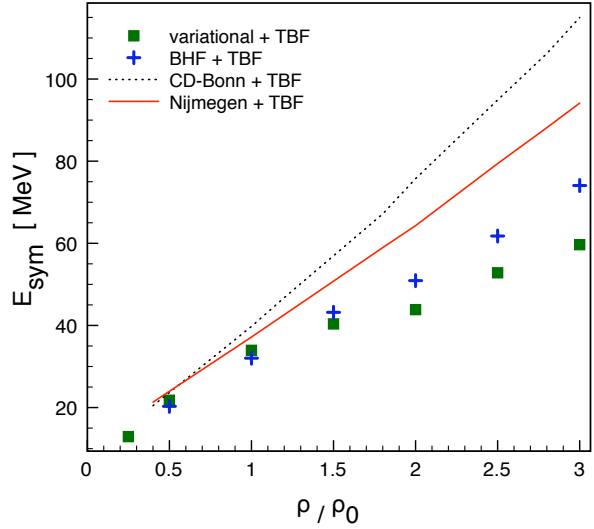}
\caption{(Color online) Symmetry energy as a function of the
density. The $T$-matrix results are compared to the variational
\cite{Akmal:1998} and BHF \cite{Baldo:2007} approaches.
}
\label{fig:symmetry}
\end{center}
\end{figure}

In Fig. \ref{fig:symmetry} our result for the symmetry energy is shown in 
comparison with the variational and BHF approaches.
When employing the CD-Bonn potential we find $S_0 = 39.7$ MeV, for Nijmegen
$S_0 = 37.1$ MeV, both slightly higher than other estimates. With 
two-body forces only we find $S_0\simeq 30 - 32$ MeV compared to $29$ MeV
from an earlier  self-consistent $T$-matrix calculation \cite{Dewulf:2003symm}. 

\begin{figure}[h]
\begin{center}
\includegraphics[width=8cm]{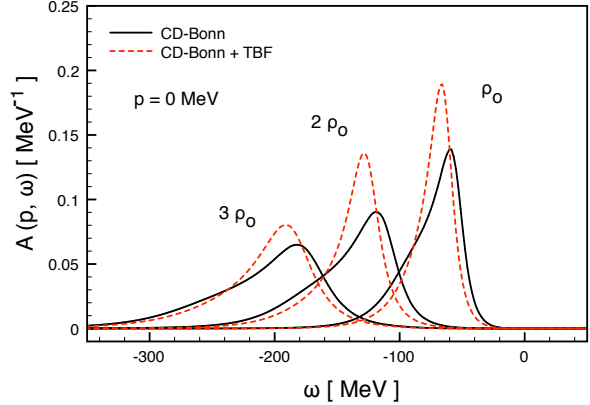}
\caption{(Color online) Spectral function in neutron matter 
at zero momentum for the CD-Bonn interaction with and without TBF.
}
\label{fig:spe-bn-0}
\end{center}
\end{figure}

Single-particle properties in neutron
matter are modified by the TBF.
 We present  the results obtained with the CD-Bonn
potential, premising that the same considerations are valid for the Nijmegen 
NN interaction.
In Fig. \ref{fig:spe-bn-0} the spectral function at zero momentum is shown for
three different densities. We see that in pure neutron matter the introduction
of three-body forces causes an enhancement of the quasi-particle peak of the
spectral function at all densities $\rho>\rho_0$.
 The suppression of correlations is more
evident at low densities and decreases as we go up to $3 \, \rho_0$.

\begin{figure}[h]
\begin{center}
\includegraphics[width=12cm]{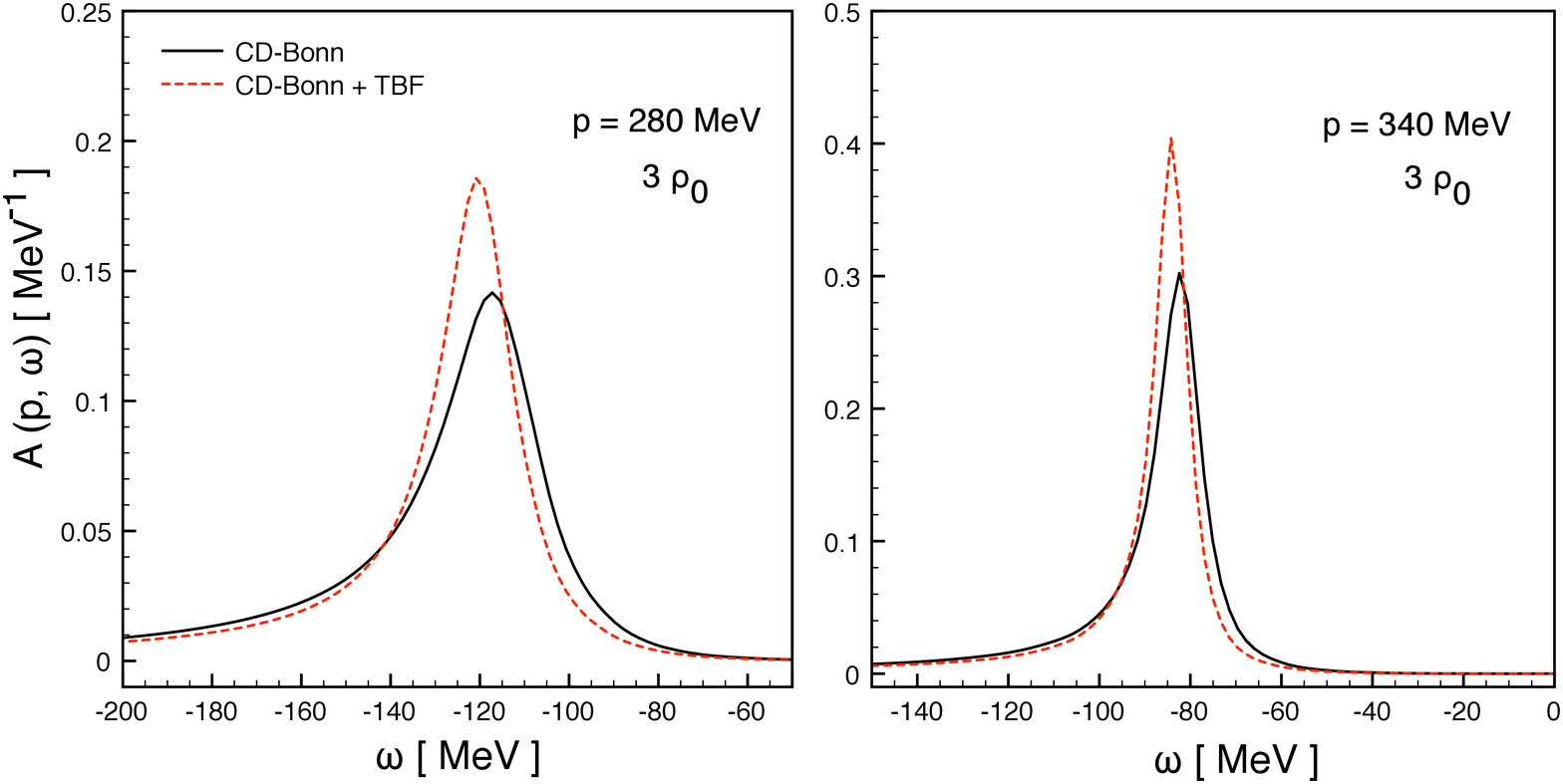}
\caption{(Color online) Nucleon spectral function in neutron matter at $ 3
  \rho_0$ for $p=280$ MeV (left panel)
and $p=340$ MeV (right panel).
}
\label{fig:spe-bn-comb}
\end{center}
\end{figure}
For all momenta up to $p_F$ there is a narrowing of the peak. The 
enhancement of the quasiparticle strength can be quantified as $20-30 \%$. 
In Fig. \ref{fig:spe-bn-comb} the
spectral functions for two other  momenta are displayed with qualitatively
 similar effects.

\begin{figure}[h]
\begin{center}
\includegraphics[width=12cm]{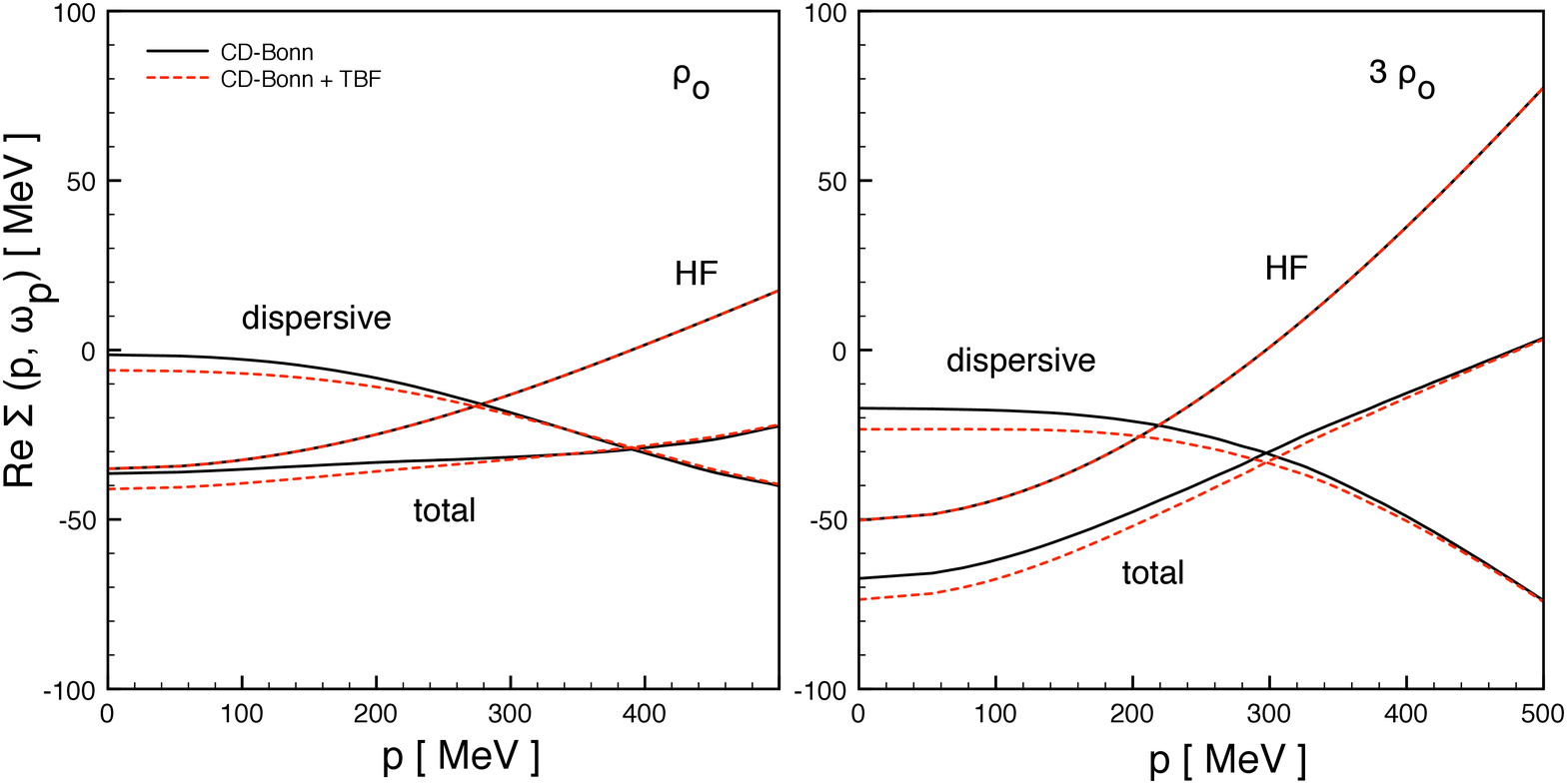}
\caption{(Color online) Different contributions to the self-energy 
$\mbox{Re} \, \Sigma (\bm{p}, \omega)$ at $\rho_0$ and $3 \, \rho_0$ in 
 neutron matter (CD-Bonn interaction).
}
\label{fig:self-bn-comb}
\end{center}
\end{figure}

The suppression of the effects of TBF above the Fermi surface is evident when
we consider the self-energy (\ref{eq:re-sigma}), plotted as a function of 
momentum in Fig. \ref{fig:self-bn-comb} for $\rho=\rho_0$ (which corresponds 
to $p_F=331$ MeV) and $\rho=3 \,\rho_0$ ($p_F=477$ MeV).
\begin{figure}[h]
\begin{center}
\includegraphics[width=12cm]{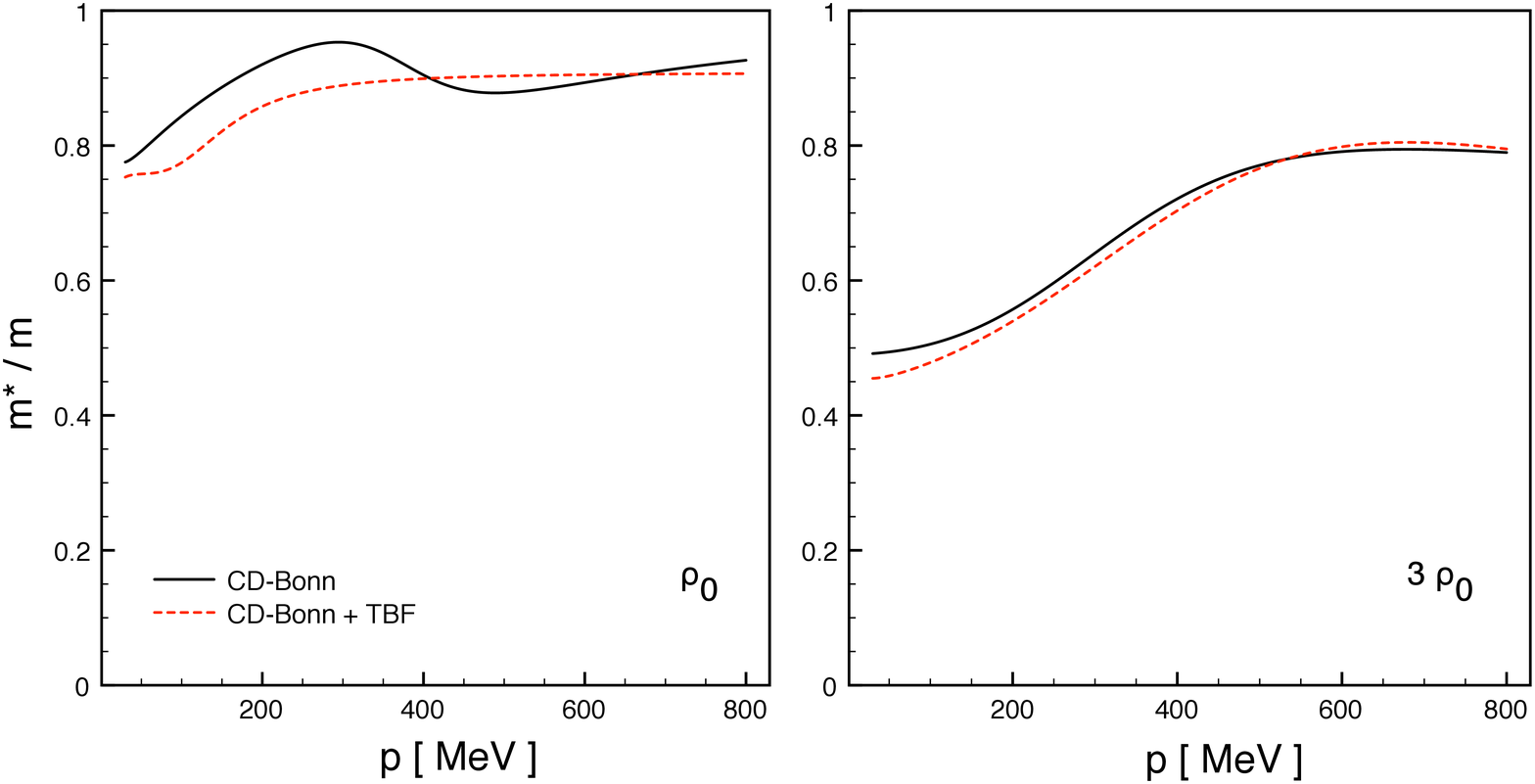}
\caption{(Color online) Effective mass in neutron matter
 (in units of the nucleon mass $m=939$ MeV) for the 
CD-Bonn interaction 
 at $\rho_0$ (left panel) and $3 \, \rho_0$ (right panel).
}
\label{fig:effm-bn}
\end{center}
\end{figure}
The self-energy is shifted down by $5-10$ MeV by the TBF, i.e. the
single-particle potential becomes slightly more attractive. This causes a
reduction of the neutron effective mass (see Fig. \ref{fig:effm-bn}) for
momenta below $p_F$. Similarly to the symmetric case, for $\rho_0$ after the
introduction of TBF the structure 
 around the Fermi momentum disappears, and the
value at the Fermi surface is shifted from $m^* \approx 0.9$ to 
$m^* \approx 0.85$ . At $3 \, \rho_0$ there are only 
small differences when including TBF. With or without TBF 
 the effective mass at the Fermi surface is found to be  $m^* \approx 0.75$.

\section{Summary and conclusions}
\label{sec:conclusions}

The equation of state of symmetric and asymmetric nuclear matter is of crucial
interest in the study of heavy ion collisions and the modeling of
neutron stars. The self-consistent Green's function technique, in which the
strong correlations induced by the nuclear interactions are resummed in the
in-medium $T$-matrix, accounts for the 
modified particle properties in the dense medium
and provides consistently the macroscopic thermodynamic observables.

Within this approach, we calculate the properties of
isospin symmetric and asymmetric nuclear matter at zero temperature, 
focusing on the impact of
three-body forces on the density dependence of the energy 
per particle and of  the single-particle
properties. We employ two different NN potentials, CD-Bonn and
Nijmegen, and the Urbana semi-phenomenological TBF. The results are
qualitatively independent of the starting NN potential, so that the present
conclusions are valid for both of them.
\begin{table}[htbp] 
\begin{center} 
\begin{tabular}{|c||c|c|c|c|c|} 
\hline   & $ \rho_0 $ & 
$ E_0 $  &
$ K_0 $  & 
$ S_0 $  & 
$ m^*/m $
\\ &  $[\mbox{fm}^{-3}] $ &  [MeV] &  [MeV] & [MeV]& \\ 
 \hline 
\hline experiment  & $0.16 \pm 0.01$ & $-16 \pm 1$ & 
$210 \pm 30$ & $32 \pm 6$  & $\approx 0.8$ 
\\ 
\hline CD-Bonn  & $0.287$ & $-19.9$ & $70$ & 
$32.2$  & $0.90$ 
\\ 
\hline CD-Bonn + TBF & $0.171$ & $-16.3$ &  $148$ & 
$39.7$  & $0.87$ 
\\ 
\hline Nijmegen & $0.235$ & $-18.4$ & $76$ & 
$30.5$  & $0.87$ 
\\
\hline Nijmegen + TBF & $0.164$ & $-16.4$ & $158$ & 
$37.1$  & $0.90$ 
\\
\hline
\end{tabular} 
\end{center} 
\caption{Summary of the main features of the nuclear matter EOS obtained with
  and without TBF, compared with experimental estimates taken from
  \cite{Haensel:NS}.
}
\label{tab:summary} 
\end{table}
In table \ref{tab:summary} we summarize the most important parameters of the 
equation of
state calculated with and without three-body forces. There is an evident
improvement after the introduction of TBF, in particular in the description of
the saturation properties of symmetric nuclear matter. The incompressibility
at saturation $K_0$ is perhaps the only unsatisfactory result, being smaller
than the current  estimates. We have tested also a stiffer set of
TBF parameters but we are not able to increase $K_0$ without spoiling the
neutron matter equation of state, which has already a stiff slope.

In general, in a dense medium three-nucleon forces play a significant role. 
The modifications
of the EOS and of the single-particle properties are larger when the 
density increases.
Analyzing the momentum dependence we notice that the effects of TBF are
maximal around the Fermi surface, and are strongly suppressed for high momenta.
 
At high densities the nucleon spectral function is modified
significantly. This can change the estimate of the in-medium
nucleon-nucleon cross section, which is relevant for the description of
heavy-ion reactions and the cooling of neutron stars.
The effects depend on the isospin asymmetry of the system. It is 
 a broadening of the quasi-particle peak of
the spectral function in the case of symmetric matter, which reflects the
increase of scattering between particles. On the contrary for neutron matter
the peak is enhanced suggesting a stronger quasiparticle behavior. 
At subsaturation densities, relevant for finite nuclei, the effect of TBF on
the spectral function is small. Thus, TBF cannot explain the observed 
strong dependence
 of the proton spectral functions on the average density and
 proton to neutron ratio in nuclei \cite{Rohe}.

We  also analyze the dispersive self-energy and the resulting optical
potential is more repulsive (attractive) with TBF in symmetric (neutron) matter. 
The effective mass is not very sensitive to the effects of TBF.
Its momentum dependence is slightly smoothed out with the 
introduction of TBF.

All the present results are at zero temperature.
Within the $T$-matrix approach, it  is possible to investigate
the thermodynamic properties of such systems at finite temperature,
including TBF forces into the  calculations. TBF are a necessary ingredient in
order to reach a realistic 
EOS of dense and hot nuclear matter.

\bibliography{vibi}

\end{document}